\documentstyle[epsf,prc,aps,graphicx]{revtex}
\begin{document}
\preprint{Saclay-T99/139}
\draft
\title{Effects of HBT correlations on flow measurements}
\author{Phuong Mai DINH$^1$, Nicolas BORGHINI$^2$ and Jean-Yves OLLITRAULT$^1$}
\address{$^1$~Service de Physique Th\'eorique, CE-Saclay\\
F-91191 Gif-sur-Yvette cedex}
\address{$^2$~Laboratoire de Physique Th\'eorique des Particules El\'ementaires 
\\ 
Universit\'e Pierre et Marie Curie\\
4, place Jussieu\\
F-75252 Paris cedex 05}
\maketitle
\begin{abstract}
The methods currently used to measure collective flow in nucleus--nucleus 
collisions assume that the only azimuthal correlations between particles 
are those arising from their correlation with the reaction plane. 
However, quantum HBT correlations also produce short range 
azimuthal correlations between identical particles. This creates 
apparent azimuthal anisotropies of a few percent when pions are used 
to estimate the direction of the reaction plane. These should not 
be misinterpreted as originating from collective flow. 
In particular, we show that 
the peculiar behaviour of the directed and 
elliptic flow of pions observed by NA49 at low $p_T$ can be entirely 
understood in terms of HBT correlations. 
Such correlations also produce apparent higher Fourier harmonics 
(of order $n\ge 3$) of the azimuthal distribution, with magnitudes 
of the order of 1\%, which should be looked for in the data.  
\end{abstract}

\section{Introduction}

In a heavy ion collision, the azimuthal distribution of particles 
with respect to the direction of impact (reaction plane) is not 
isotropic for non-central collisions. 
This phenomenon, referred to as collective flow, was first 
observed fifteen years ago at Bevalac~\cite{Gust84}, and 
more recently at the higher AGS~\cite{Barr94} and SPS~\cite{Wien96}
energies. 
Azimuthal anisotropies are very sensitive to nuclear matter
properties~\cite{DANI4,Sorge}. It is therefore important to measure
them accurately. 
Throughout this paper, we use the word ``flow'' in the restricted 
meaning of ``azimuthal correlation between the directions of outgoing 
particles and the reaction plane''. We do not 
consider radial flow~\cite{radial}, which is usually measured for 
central collisions only. 

Flow measurements are done in three steps
(see \cite{VOLO/POSK} for a recent review of the methods):
first, one estimates the direction of the reaction plane event by event
from the directions of the outgoing particles;
then, one measures the azimuthal distribution of particles with 
respect to this estimated reaction plane; finally, one  
corrects this distribution for the statistical error in the 
reaction plane determination. 
In performing this analysis, one usually assumes that the only 
azimuthal correlations between particles result from their 
correlations with the reaction plane, i.e. from flow 
This implicit assumption is made, in particular, in the ``subevent'' method 
proposed by Danielewicz and Odyniec~\cite{DANI} in order to estimate 
the error in the reaction plane determination. This method 
is now used by most, if not all, heavy ion experiments. 

However, other sources of azimuthal correlations are known, 
which do not depend on the orientation of the reaction plane. 
For instance, there are quantum correlations between identical 
particles, due to the (anti)symmetry of the wave function~: 
this is the so-called Hanbury-Brown and Twiss effect~\cite{HBT}, 
hereafter denoted by HBT (see \cite{BOAL,WIEDE/HEINZ} for reviews).
Azimuthal correlations due to the HBT effect have been studied 
recently in~\cite{Mrow}. 
In the present paper, we show that if the standard flow analysis is
performed, these correlations produce a spurious flow. 
This effect is important when pions are used to estimated 
the reaction plane, which is often the case at ultrarelativistic 
energies, in particular for the NA49 experiment at CERN~\cite{Data}. 
We show that when these correlations are properly subtracted, the 
flow observables are considerably modified at low transverse momentum. 

In section 2, we recall how the Fourier coefficients of the 
azimuthal distribution with respect to the reaction plane 
are extracted from the two-particle correlation function in the 
standard flow analysis. 
Then, in section 3, we apply this procedure to the measured 
two-particle HBT correlations, and calculate the spurious 
flow arising from these correlations. 
Finally, in section 4, we explain how to subtract HBT 
correlations in the flow analysis, and perform this subtraction on 
the NA49 data, using the HBT correlations measured by the 
same experiment. 
Conclusions are presented in section 5. 

\section{Standard flow analysis}
\label{s:flow}

In nucleus--nucleus collisions, 
the determination of the reaction plane event by event allows in
principle to measure the distribution of particles not only in transverse
momentum $p_T$ and rapidity $y$, but also in azimuth $\phi$, where 
$\phi$ is the azimuthal angle with respect to the reaction plane.
The $\phi$ distribution is conveniently characterized by its 
Fourier coefficients~\cite{VOLO/ZH}
\begin{equation}
\label{vn}
v_n(p_T,y)\equiv\langle\cos n\phi\rangle =
{\int_0^{2\pi} \cos n\phi {\displaystyle 
dN\over\displaystyle d^3{\bf p}}d\phi\over
\int_0^{2\pi}{\displaystyle dN\over\displaystyle
 d^3{\bf p}}d\phi}
\end{equation}
where the brackets denote an average value over many events. 
Since the system is symmetric with respect to the reaction plane
for spherical nuclei, $\langle\sin n\phi\rangle$ vanishes. 
Most of the time, because of limited statistics, $v_n$ is averaged 
over $p_T$ and/or $y$. 
The average value of $v_n(p_T,y)$ 
over a domain ${\cal D}$ of the $(p_T,y)$ plane, 
corresponding to a detector, will be denoted by $v_n({\cal D})$. 
In practice, the published data are limited to the $n=1$ (directed flow) 
and $n=2$ (elliptic flow) coefficients. 
However, higher harmonics could reveal 
more detailed features of the $\phi$ distribution~\cite{VOLO/POSK}. 

Since the orientation of the reaction plane is not known a priori, 
$v_n$ must be extracted from the azimuthal correlations between 
the produced particles. 
We introduce the two-particle distribution, which is generally written as 
\begin{equation}
\label{defc2}
{dN\over d^3{\bf p_1}d^3{\bf p_2}}=
{dN\over d^3{\bf p_1}}{dN\over d^3{\bf p_2}}
\left(1+C({\bf p}_1,{\bf p}_2)\right)
\end{equation}
where $C({\bf p}_1,{\bf p}_2)$ is the two-particle connected 
correlation function, which vanishes for independent particles. 
The Fourier coefficients of the relative azimuthal distribution 
are given by 
\begin{equation}
\label{defcn}
c_n(p_{T1},y_1,p_{T2},y_2)
\equiv \langle\cos n(\phi_1-\phi_2)\rangle
={\int\!\!\!\int{\cos n(\phi_1-\phi_2) 
{\displaystyle dN\over \displaystyle d^3{\bf p}_1 d^3{\bf p}_2}d\phi_1d\phi_2}
\over\int\!\!\!\int{{\displaystyle dN\over \displaystyle 
d^3{\bf p}_1d^3{\bf p}_2}d\phi_1d\phi_2}}.
\end{equation}
We denote the average value of $c_n$  over $(p_{T2},y_2)$ 
in the domain ${\cal D}$ by $c_n(p_{T1},y_1,{\cal D})$, 
and the average over both 
$(p_{T1},y_1)$  and $(p_{T2},y_2)$ by $c_n({\cal D},{\cal D})$. 

Using the decomposition (\ref{defc2}), one can write $c_n$ as the sum 
of two terms:
\begin{equation}
\label{decomposition}
c_n(p_{T1},y_1,p_{T2},y_2)= 
c_n^{\rm flow}(p_{T1},y_1,p_{T2},y_2)+
c_n^{\rm non-flow}(p_{T1},y_1,p_{T2},y_2)
\end{equation}
where the first term is due to flow:
\begin{equation}
\label{flo1}
c_n^{\rm flow}(p_{T1},y_1,p_{T2},y_2)=
v_n(p_{T1},y_1)v_n(p_{T2},y_2)
\end{equation}
and the remaining term comes from two-particle correlations:
\begin{equation}
\label{cnnonflow}
c_n^{\rm non-flow}(p_{T1},y_1,p_{T2},y_2)=
{\int\!\!\!\int{\cos n(\phi_1-\phi_2) 
C({\bf p}_1,{\bf p}_2){\displaystyle dN\over \displaystyle d^3{\bf p}_1 }
{\displaystyle dN\over \displaystyle d^3{\bf p}_2 }d\phi_1d\phi_2}
\over\int\!\!\!\int{{\displaystyle dN\over \displaystyle 
d^3{\bf p}_1d^3{\bf p}_2}d\phi_1d\phi_2}}
\end{equation}
In writing Eq.(\ref{flo1}), we have used the fact that 
$\langle\sin n\phi_1\rangle=\langle\sin n\phi_2\rangle=0$ 
and neglected the correlation $C({\bf p}_1,{\bf p}_2)$ in the 
denominator. 

In the standard flow analysis, non-flow correlations are 
neglected~\cite{VOLO/POSK,DANI}, with a few exceptions:
the correlations due to momentum conservation are taken into 
account at intermediate energies~\cite{DANI2}, and 
correlations between photons originating from $\pi^0$ decays 
were considered in~\cite{WA93}. The effect of non-flow correlations 
on flow observables is considered from a general point of view 
in \cite{Olli}. 
In the remainder of this section, we assume that $c_n^{\rm non-flow}=0$. 
Then, $v_n$ can be calculated simply as a function of the measured 
correlation $c_n$ using Eq.(\ref{flo1}), as we now show. 
Note, however, that Eq.(\ref{flo1}) is invariant under a global change 
of sign: $v_n(p_T,y)\to -v_n(p_T,y)$. Hence the sign of $v_n$ 
cannot be determined from $c_n$.  
It is fixed either by physical considerations 
or by an independent measurement. 
For instance, NA49 chooses the minus sign for the $v_1$ of 
charged pions, in order to make the $v_1$ of protons at forward rapidities
come out positive~\cite{Data}. 
Averaging Eq.(\ref{flo1}) over $(p_{T1},y_1)$ and $(p_{T2},y_2)$ in 
the domain ${\cal D}$, one obtains~:
\begin{equation}
\label{flo2}
v_n({\cal D})=
\pm\sqrt{c_n({\cal D},{\cal D})}.
\end{equation}
This equation shows in particular that the average two-particle 
correlation $c_n({\cal D},{\cal D})$ due to flow is  positive. 
Finally, integrating (\ref{flo1}) over $(p_{T2},y_2)$, 
and using (\ref{flo2}), one obtains the expression of $v_n$ as 
a function of $c_n$:
\begin{equation}
\label{flo3}
v_n (p_{T1},y_1)=
\pm{c_n(p_{T1},y_1,{\cal D})
\over 
\sqrt{c_n({\cal D},{\cal D})}}.
\end{equation}
This formula serves as a basis for the standard flow analysis. 

Note that the actual experimental procedure is usually 
different: one first estimates, for a given Fourier harmonic $m$, the azimuth 
of the reaction plane (modulo $2\pi/m$) by summing over many particles. 
Then one studies the correlation of another 
particle (in order to remove autocorrelations) 
with respect to the estimated reaction plane. 
One can then measure the coefficient $v_n$ with respect to this 
reaction plane if $n$ is a multiple of $m$. 
In this paper, we consider only the case $n=m$. 
Both procedures give the same result, since they start from the 
same assumption (the only azimuthal correlations are from flow). 
This equivalence was first pointed out in \cite{WANG}. 

\section{Azimuthal correlations due to the HBT effect}

The HBT effect yields two-particle correlations, i.e. 
a non-zero $C({\bf p}_1,{\bf p}_2)$ in Eq.(\ref{defc2}). 
According to Eq.(\ref{cnnonflow}), this gives rise to an azimuthal 
correlation $c_n^{\rm non-flow}$, which contributes to the total, 
measured correlation $c_n$ in Eq.(\ref{decomposition}). 
In particular, there will be a correlation between randomly 
chosen subevents 
when one particle of a HBT pair goes into each subevent. 
The contribution of HBT correlations to $c_n^{\rm non-flow}$
will be denoted by $c_n^{\rm HBT}$. 

In the following, we shall consider only 
pions. Since they are bosons, their correlation is positive, 
i.e. of the same sign as the correlation due to flow. 
Therefore, if one applies the standard flow analysis to HBT 
correlations alone, i.e. if one replaces $c_n$ by $c_n^{\rm HBT}$ in 
Eq.(\ref{flo3}), they yield a spurious flow $v_n^{\rm HBT}$, 
which we calculate in this section. 

First, let us estimate its order of magnitude. 
The HBT effect gives a correlation of order unity between 
two identical pions with momenta ${\bf p_1}$ and ${\bf p_2}$ 
if $|{\bf p_2}-{\bf p_1}|\lesssim \hbar/R$, where $R$ is a typical 
HBT radius, corresponding to the size of the interaction 
region. From now on, we take $\hbar=1$. 
In practice, $R\sim 4$~fm for a semi--peripheral Pb--Pb collision at 
158~GeV per nucleon, so that 
$1/R\sim 50$~MeV/c is much smaller than the average transverse 
momentum, which is close to $400$~MeV/c: the HBT effect 
correlates only pairs with low relative momenta. 

In particular, the azimuthal correlation due to the HBT effect 
is short-ranged~: it is significant only if 
$\phi_2-\phi_1\lesssim 1/(Rp_T)\sim 0.1$.  
This localization in $\phi$ implies a delocalization in $n$ 
of the Fourier coefficients, which are expected to be roughly 
constant up to $n\lesssim Rp_T\sim 10$, as will be confirmed below. 

For small $n$ and $(p_{T1},y_1)$ in  ${\cal D}$, 
the order of magnitude of $c_n^{\rm HBT}(p_{T1},y_1,{\cal D})$ 
is the fraction of particles in ${\cal D}$ whose momentum 
lies in a circle of radius $1/R$ centered at ${\bf p_1}$. 
This fraction is of order 
$(R^3 \langle p_T\rangle^2 \langle m_T\rangle\Delta y)^{-1}$, 
where $\langle p_T\rangle$ and $\langle m_T\rangle$ are typical 
magnitudes of the transverse momentum and transverse mass
($m_T=\sqrt{p_T^2+m^2}$, where $m$ is the mass of the particle), 
respectively, while $\Delta y$ is the rapidity interval covered 
by the detector. Using Eq.(\ref{flo2}), this gives a spurious 
flow of order 
\begin{equation}
\label{orderofmagnitude}
\left| v_n^{\rm HBT}({\cal D})\right|
\sim \left({1\over R^3 \langle p_T\rangle^2 
\langle m_T\rangle\Delta y}\right)^{1/2}.
\end{equation}
The effect is therefore larger for the lightest particles, i.e. for 
pions. 
Taking $R=4$~fm, $\langle p_T\rangle\sim \langle m_T\rangle\sim
400$~MeV/c  
and $\Delta y=2$, one obtains $\left|v_n({\cal D})\right|\sim 3$~\%, 
which is of the same order 
of magnitude as the flow values measured at SPS.  
It is therefore {\it a priori} important to take HBT correlations 
into account in the flow analysis. 

We shall now turn to a more quantitative estimate of $c_n^{\rm HBT}$. 
For this purpose, we use the standard gaussian parametrization of 
the correlation function (\ref{defc2}) 
between two identical pions\cite{LCMS}:
\begin{equation}
\label{c2}
C({\bf p}_1,{\bf p}_2)=\lambda 
e^{-q_s^2 R_s^2-q_o^2 R_o^2-q_L^2 R_L^2}
\end{equation}
One chooses a frame boosted along the collision axis in such a way 
that $p_{1z}+p_{2z}=0$ (``longitudinal comoving system'', denoted by 
LCMS). 
In this frame, $q_L$, $q_o$ and $q_s$ denote the projections of 
${\bf p}_2-{\bf p}_1$ along the collision axis, 
the direction of ${\bf p}_1+{\bf p}_2$ and the third direction, respectively. 
The corresponding radii $R_L$, $R_o$ and $R_s$, as well as the 
parameter $\lambda$ ($0\le\lambda\le 1$), depend on 
${\bf p}_1+{\bf p}_2$. We neglect this dependence
in the following calculation. 
Note that the parametrization (\ref{c2}) is valid for central collisions, 
for which the pion source is azimuthally symmetric. Therefore the 
azimuthal correlations studied in this section have nothing to do with flow. 
Note also that we neglect Coulomb correlations, which should be taken 
into account in a more careful study. We hope that repulsive Coulomb
correlations between like-sign pairs will be compensated, at least 
partially, by attractive correlations between opposite sign pairs. 

Since $C({\bf p}_1,{\bf p}_2)$ vanishes unless 
${\bf p}_2$ is very close to ${\bf p}_1$, we may replace 
$dN/d^3{\bf p_2}$ by $dN/d^3{\bf p_1}$ in the numerator  of 
Eq.(\ref{cnnonflow}), and then integrate over ${\bf p}_2$. 
As we have already said, $q_s$, $q_o$ and $q_L$ are 
the components of ${\bf p}_2-{\bf p}_1$ in the LCMS, and one can 
equivalently integrate over $q_s$, $q_o$ and $q_L$. 
In this frame, $y_1\simeq 0$ and one may also replace 
$dN/d^3{\bf p}_1$ by $(1/m_{T1})dN/d^2{\bf p_T}_1 dy_1$. 
The resulting formula is boost invariant and can also be used in the 
laboratory frame. 

The relative angle $\phi_2-\phi_1$ can be expressed as 
a function of $q_s$ and $q_o$. 
If $p_{T1}\gg 1/R$, 
then to a good approximation 
\begin{equation}
\label{phiq}
\phi_2-\phi_1\simeq {q_s/p_{T1}}.
\end{equation}
If $p_{T1}\sim 1/R$, Eq.(\ref{phiq}) 
is no longer valid. We assume that $R_s\simeq R_o$ and use, instead of 
(\ref{phiq}), the following relation~:
\begin{equation}
\label{phi12}
q_s^2+q_o^2=p_{T1}^2+p_{T2}^2-2 p_{T1}p_{T2}\cos(\phi_2-\phi_1). 
\end{equation}

To calculate $c_n^{\rm HBT}(p_{T1},y_1,{\cal D})$, we insert 
Eqs.(\ref{c2}) and (\ref{phiq}) in the numerator of (\ref{cnnonflow})
and integrate over $(q_s,q_o,q_L)$. 
The limits on $q_o$ and $q_L$ 
are deduced from the limits on $(p_{T2},y_2)$, using the following
relations, valid if $p_{T1}\gg 1/R$~:
\begin{eqnarray}
q_o&=&p_{T2}-p_{T1}\cr
q_L&=&m_{T1}(y_2-y_1).
\end{eqnarray}
Since $q_s$ is independent of $p_{T2}$ and $y_2$ (see Eq.(\ref{phiq})),
the integral over $q_s$ extends from $-\infty$ to $+\infty$.

Note that values of $q_o$ and $q_L$ much larger than $1/R$ do not 
contribute to the correlation (\ref{c2}), 
so that one can extend the integrals 
over $q_o$ and $q_L$ to $\pm\infty$ as soon as the point $(p_{T1},y_1)$
lies in ${\cal D}$ and is not too close to the boundary of ${\cal
D}$. By too close, we mean 
within an interval $1/R_o\sim 50$~MeV/c in $p_T$ or 
$1/(R_L m_T)\sim 0.3$ in $y$. One then obtains after integration
\begin{equation}
\label{corr2}
c_n^{\rm HBT}(p_{T1},y_1,{\cal D})=
{\lambda\pi^{3/2}\over R_sR_oR_L}
\exp\left(-{n^2\over 4 p_{T1}^2 R_s^2}\right)
{\displaystyle{1\over m_{T1}}{dN\over d^2 {\bf p_T}_1 dy_1}\over 
\displaystyle\int_{\cal D}{dN\over d^2 {\bf p_T}_2 dy_2}
d^2 {\bf p_T}_2 dy_2}.
\end{equation} 
At low $p_T$, Eq.(\ref{phiq}) must be replaced by Eq.(\ref{phi12}). 
Then, one must do the following substitution in Eq.(\ref{corr2})~:
\begin{equation}
\exp\left(-{n^2\over 4\chi^2}\right)\rightarrow
{\sqrt{\pi}\over 2}\chi e^{-\chi^2/2}\left(
I_{n-1\over 2}\left( {\chi^2\over 2}\right) +
I_{n+1\over 2}\left( {\chi^2\over 2}\right) \right)
\end{equation}
where $\chi=R_s p_T$ and $I_k$ is the modified Bessel function of order $k$.

Let us discuss our result (\ref{corr2}).  
First, the correlation depends on $n$ 
only through the exponential factor, which suppresses $c_n^{\rm HBT}$ 
in the very low $p_T$ region $p_{T1}\lesssim n/2 R_s$. 
For $n$ smaller than $R_s\langle p_T\rangle\simeq 10$, 
the correlation depends weakly on $n$, as discussed above. 
Neglecting this $n$ dependence, (\ref{corr2}) reproduces the order 
of magnitude (\ref{orderofmagnitude}). To see this, we normalize the 
particle distribution in ${\cal D}$ in order to get rid of the denominator 
in (\ref{corr2}), and the numerator 
$(1/ m_{T1})(dN/d^2{\bf p_T}_1 dy_1)$
is of order $1/\langle p_T\rangle^2 \langle m_T\rangle\Delta y$. 
However, Eq.(\ref{corr2}) is more detailed, and shows in particular 
that the dependence of the correlation on $p_{T1}$ and $y_1$ 
follows that of the momentum distribution in the LCMS 
(neglecting the $m_T$ and $y$ dependence of HBT radii). 
This is because the correlation $c_n^{\rm HBT}$ is proportional to 
the number of particles surrounding ${\bf p}_1$ in phase space. 

Let us now present numerical estimates for a Pb--Pb collision at 
SPS. 
We assume for simplicity that the $p_T$ and $y$ dependence of the 
particle distribution factorize, thereby neglecting 
the observed variation of $\langle p_T\rangle$ 
with rapidity~\cite{Spectra}.
The rapidity dependence of charged pions 
can be parametrized by~\cite{Spectra}:
\begin{equation}
\label{yspectrum}
{dN\over dy}= \frac{1}{\sigma \sqrt{2\pi}}
	      \exp \left( -\frac{\left(y-\langle y\rangle\right)^2}
			       {2 \sigma^2} \right)
\end{equation}
with $\sigma = 1.4$ and $\langle y\rangle=2.9$.
The  normalized $p_T$ distribution is parametrized by 
\begin{equation}
\label{mtspectrum}
{dN\over d^2{\bf p_T}}= {e^{m/T}\over 2\pi T (m+T)}\exp\left(-{m_T\over
T}\right).
\end{equation}
with $T\simeq 190$~MeV~\cite{Spectra}. 
This parametrization underestimates the number of low-$p_T$
pions. 
The values of $R_o$, $R_s$ and $R_L$ used in our computations, 
taking into account that the collisions are semi-peripheral, 
are respectively 4 fm, 4 fm and 5 fm~\cite{Rayons HBT}.
The correlation strength $\lambda$ is 
approximately 0.4 for pions \cite{Seyboth}.

Finally, we must define the domain ${\cal D}$ in Eq.(\ref{corr2}). 
It is natural to choose different rapidity windows for odd and even 
harmonics, because 
odd harmonics have opposite signs in the target and projectile 
rapidity region, by symmetry, and vanish at mid-rapidity 
($\langle y\rangle=2.9$), while even harmonics are symmetric around
mid-rapidity. 
Following the NA49 collaboration~\cite{private}, we take 
$4<y<6$ and $0.05<p_T<0.6$~GeV/c for odd $n$, and 
$3.5<y<5$ and $0.05<p_T<2$~GeV/c for even $n$. 
We assume that the particles in ${\cal D}$ are 85\% pions~\cite{Data}, 
half $\pi^+$, half $\pi^-$. 
Then, for an identified charged pion (a $\pi^+$, say) with $p_T=p_{T1}$ 
and $y=y_1$, the right-hand side of Eq.(\ref{corr2}) must be multiplied 
by $0.85\times 0.5$, which is the probability that a particle in ${\cal D}$
be also a $\pi^+$.

\begin{figure}[htb]
\begin{center}
\includegraphics[width=0.47\linewidth]{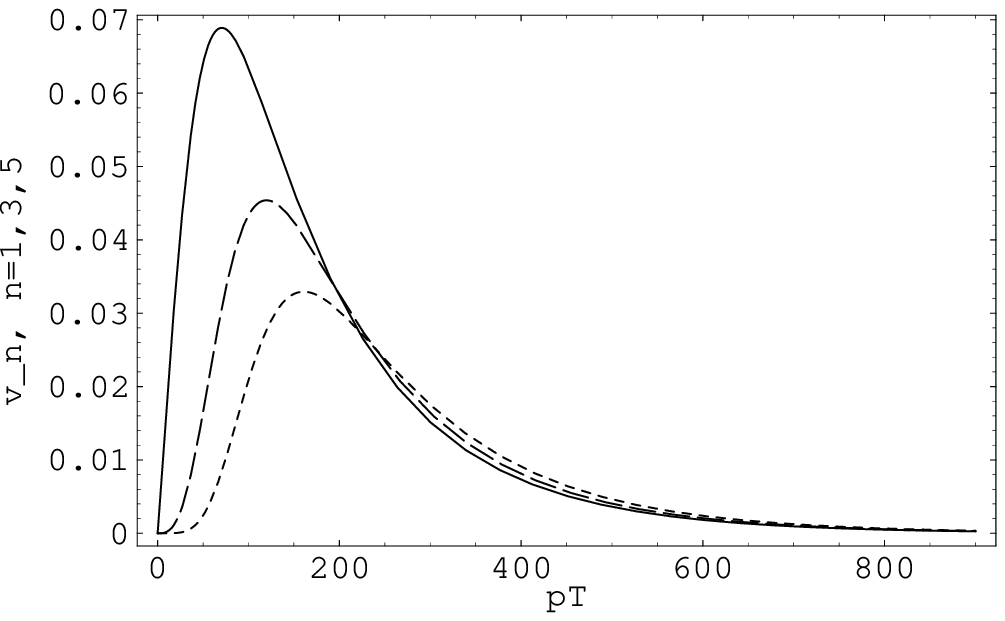}
\hspace{.5cm}
\includegraphics[width=0.47\linewidth]{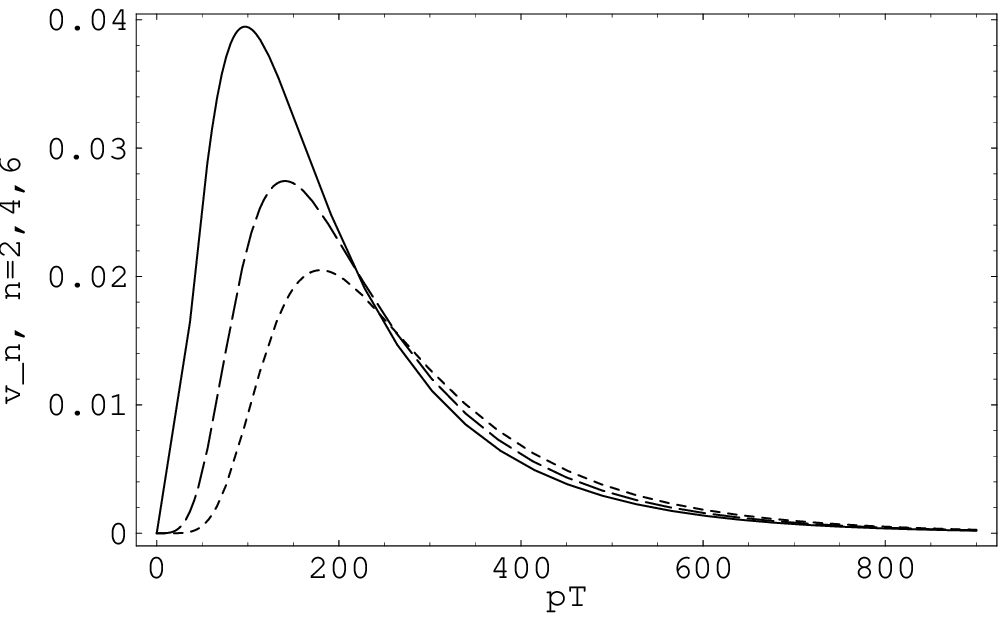}
\end{center}
\caption{Apparent $|v_n^{\rm HBT}|$ from HBT correlations as a function 
of $p_T$ in MeV/c. Top~: $n=1$ (solid curve), $n=3$ (long dashes), 
$n=5$ (short dashes). 
Bottom~: $n=2$ (solid curve), $n=4$ (long dashes), $n=6$ (short dashes).
The values are smaller for even $n$ than for odd $n$ 
because the rapidity intervals chosen to estimate the 
reaction plane differ. }
\label{fig:fig1}
\end{figure}

Substituting the correlation calculated from Eq.(\ref{corr2}) in 
Eq.(\ref{flo3}), one obtains the value of the spurious flow 
$v_n^{\rm HBT}(p_T,y)$ due to the HBT effect. 
Fig.\ref{fig:fig1} displays $\left|v_n^{\rm HBT}\right|$, 
integrated between $4<y<5$ (as are the NA49 data) as a function of 
$p_T$. 
As expected, $v_n^{\rm HBT}$ depends on the order $n$ only at low $p_T$, 
where it vanishes due to the exponential factor in Eq.(\ref{corr2}).  
HBT correlations, which follow the 
momentum distribution, also vanish if $p_T$ is much 
larger than the average transverse momentum. 
Assuming that $1/R_s\ll m,T$, we find from Eq.(\ref{corr2}) 
that the correlation is maximum at 
$p_T=p_{T\max}$ where
\begin{equation}
\label{ptmax}
p_{T\max}=\left( {2T\over m+T}\right)^{1/4}\sqrt{nm\over 2 R_s}\simeq
60\sqrt{n}~{\rm MeV/c},
\end{equation}
which reproduces approximately the maxima in Fig.\ref{fig:fig1}. 

Although data on higher order harmonics are still unpublished, they 
were shown at the Quark Matter '99 conference by the NA45 
Collaboration~\cite{NA45} which reports values of $v_3$ and $v_4$ 
of the same order as $v_1$ and $v_2$, respectively, suggesting that 
most of the effect is due to HBT correlations. 
Similar results were found with NA49 data~\cite{higher}.

\section{Subtraction of HBT correlations}
\label{s:subtraction}

Now that we have evaluated the contribution of HBT correlations 
to $c_n^{\rm non-flow}$, we can subtract this term from 
the measured correlation (left-hand side of Eq.(\ref{decomposition}),
which will be denoted by $c_n^{\rm measured}$ in this section) to 
isolate the correlation due to flow. 
Then, the flow $v_n$ can be calculated using Eq.(\ref{flo3}), 
replacing in this equation $c_n$ 
by the corrected correlation 
$c_n^{\rm flow}=c_n^{\rm measured}-c_n^{\rm HBT}$. 
In this section, we show the result of this modification on the 
directed and elliptic flow data published by NA49 for pions~\cite{Data}. 

The published data do not give directly the two-particle correlation 
$c_n^{\rm measured}$, but rather the measured flow $v_n^{\rm measured}$. 
Since these analyses assume that the correlation factorizes according 
to Eq.(\ref{flo1}), we can reconstruct the measured correlation as
a function of the measured $v_n$. In particular, 
\begin{equation}
c_n^{\rm measured}(p_{T1},y_1,{\cal D})=
v_n^{\rm measured} (p_{T1},y_1)v_n^{\rm measured}({\cal D}). 
\end{equation}
We then perform the subtraction of HBT correlations in both the 
numerator and the denominator of Eq.(\ref{flo3}). 

The integrated flow values measured by NA49 are 
$v_1^{\rm measured}({\cal D})=-3.0\pm 0.1\%$ 
and $v_2^{\rm measured}({\cal D})=3.0\pm 0.1\%$~\cite{private}. 
After subtraction of HBT correlations, the 
coefficients are smaller by some 20\%~:
$v_1({\cal D})=-2.5\%$ and 
$v_2({\cal D})=2.6\%$. 

\begin{figure}[htb]
\begin{center}
\includegraphics[width=0.47\linewidth]{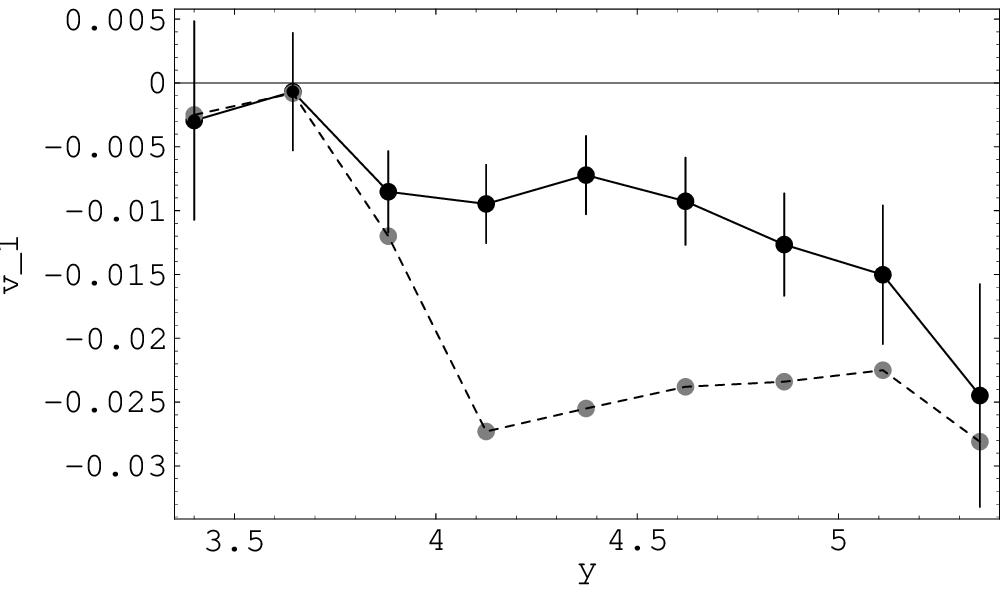}
\hspace{.5cm}
\includegraphics[width=0.47\linewidth]{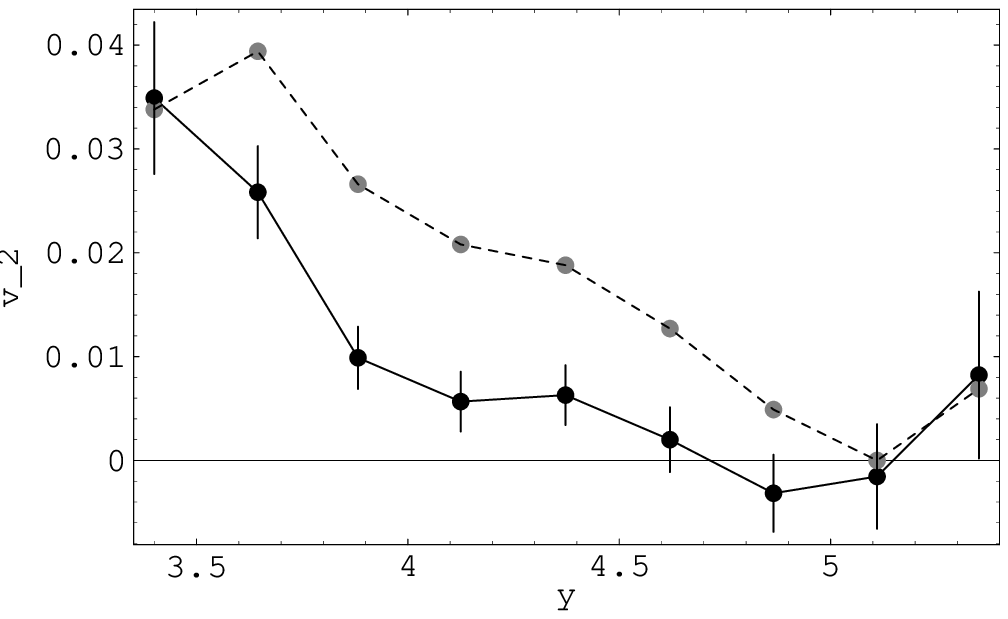}
\end{center}
\caption{Directed flow $v_1$ and elliptic flow $v_2$ of pions, 
integrated over $50<p_T<350$~MeV/c,
as a function of the rapidity, 
measured by NA49 ($v_n^{\rm measured}$, dashed curves) 
and after subtraction of HBT correlations ($v_n$, full curves). 
For clarity sake, experimental 
error bars are indicated for the corrected data only.} 
\label{fig:fig2}
\end{figure}

Fig.\ref{fig:fig2} displays the rapidity dependence of $v_1$ and 
$v_2$ at low transverse momentum, where the effect of HBT correlations 
is largest. 
Let us first comment on the uncorrected data. 
We note that $v_1^{\rm measured}$ is zero below $y<4$ 
(i.e. outside ${\cal D}$, 
where there are no HBT correlations) 
and jumps to a roughly constant value when $y>4$ 
(where HBT correlations set in). This gap disappears once HBT 
correlations are subtracted, and the resulting values of $v_1$ 
are considerably smaller. The values of $v_2$ are also much 
smaller after correction, except near mid-rapidity. 

\begin{figure}[htb]
\begin{center}
\includegraphics[width=0.47\linewidth]{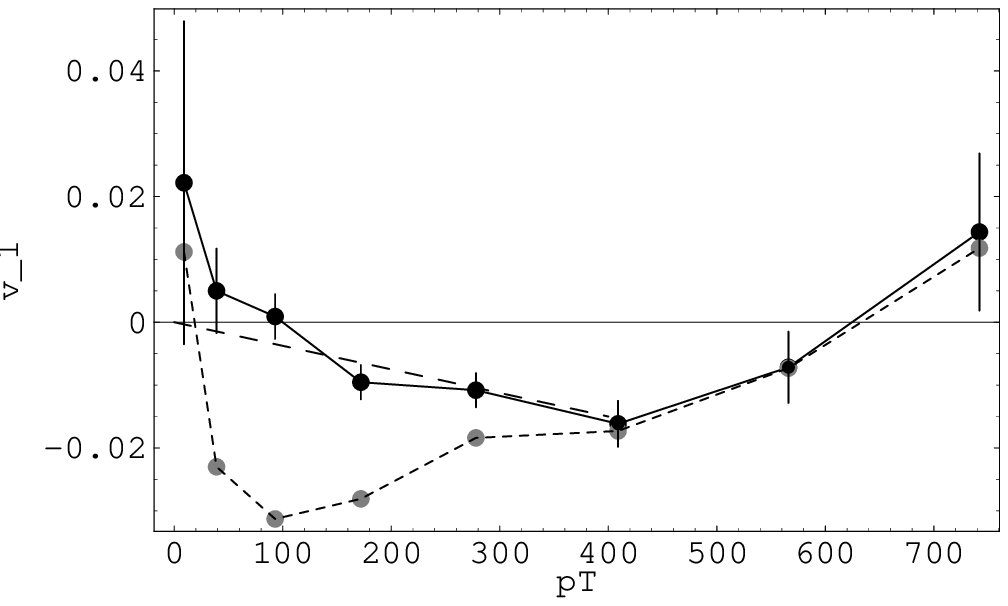}
\hspace{.5cm}
\includegraphics[width=0.47\linewidth]{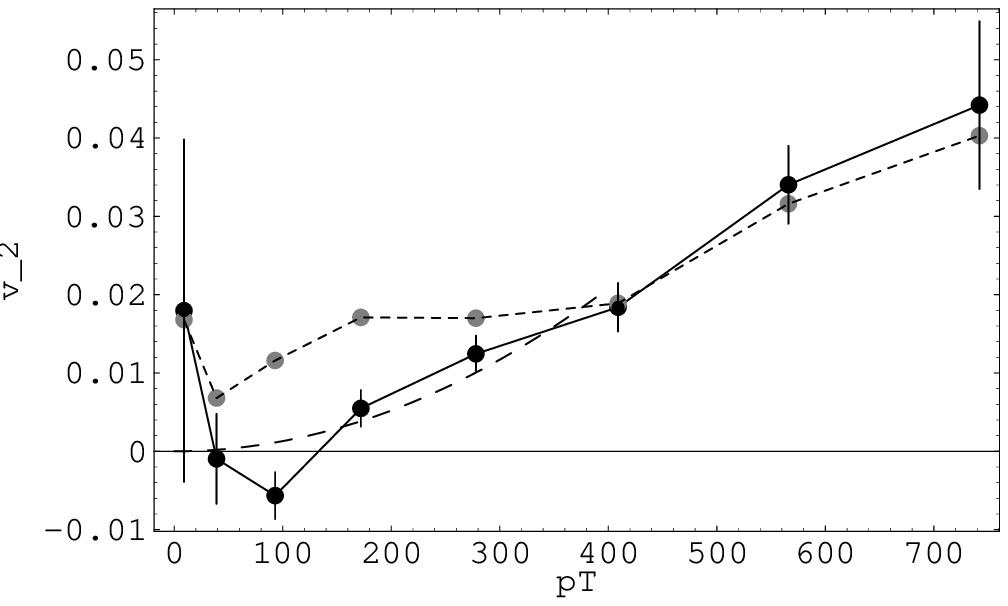}
\end{center}
\caption{Directed flow $v_1$ and elliptic flow $v_2$ of pions, 
integrated between $4<y<5$, 
as a function of the transverse momentum $p_T$ in MeV/c, 
measured by NA49 ($v_n^{\rm measured}$, short dashes) 
and after subtraction of HBT correlations ($v_n$, full curves). 
The long dashes show linear and quadratic fits at low $p_T$ for 
$v_1$ and $v_2$, respectively.} 
\label{fig:fig3}
\end{figure}

Fig.\ref{fig:fig3} displays the $p_T$ dependence of $v_1$ and $v_2$. 
The behaviour of $v_n(p_T)$ is constrained at low $p_T$~:
if the momentum distribution is regular at ${\bf p_T}={\bf 0}$, 
then $v_n(p_T)$ must vanish like $p_T^n$. 
One naturally expects this decrease to occur on a scale of 
the order of the average $p_T$. This is what is observed 
for protons~\cite{Data}. However, the uncorrected 
$v_1^{\rm measured}$ and $v_2^{\rm measured}$ 
for pions remain large far below 400~MeV/c. 
In order to explain this behaviour, one would need to invoke a specific 
phenomenon occurring at low $p_T$. No such phenomenon is known. 
Even though resonance (mostly $\Delta$) decays are known to populate 
the low-$p_T$ pion spectrum, they are not expected to produce 
any spectacular increase in the flow. 

HBT correlations provide this low-$p_T$ scale, since they are 
important down to $1/R\simeq 50$~MeV/c. 
Once they are subtracted, the peculiar behaviour of the pion 
flow at low $p_T$ disappears. $v_1$ and $v_2$ are now compatible 
with a variation of the type $v_1\propto p_T$ and $v_2\propto p_T^2$, 
up to $400$~MeV/c.

\section{Conclusions} 

We have shown that the HBT effect produces correlations
which can be misinterpreted as 
flow when pions are used to estimate the reaction plane. 
This effect is present only for pions, in the $(p_T,y)$ window used to 
estimate the reaction plane.
%, and if the Fourier harmonic 
%under study ($v_n$) is the same as that used to estimate
%the reaction plane. 
Azimuthal correlations due to the HBT effect 
depend on $p_T$ and $y$ like the momentum 
distribution in the LCMS, i.e. $(1/m_T)dN/dyd^2p_T$, and depend weakly 
on the order of the harmonic $n$. 

The pion flow observed by NA49 has peculiar features at low $p_T$: 
the rapidity dependence of $v_1$ 
is irregular, and both $v_1$ and $v_2$ remain large down 
to values of $p_T$ much smaller than the average transverse 
momentum, while they should decrease with $p_T$ as $p_T$ and $p_T^2$, 
respectively. 
All these features disappear once HBT correlations are properly 
taken into account. 
Furthermore, we predict that HBT correlations should also produce 
spurious higher harmonics of the pion azimuthal distribution 
($v_n$ with $n\ge 3$) at low $p_T$, weakly decreasing with 
$n$, with an average value of the order of 1\%. 
The data on these higher harmonics should be published. 
This would provide a confirmation of the role played by HBT correlations. 
More generally, our study shows that although non-flow azimuthal 
correlations are neglected in most analyses, they may be significant. 
\bigskip

{\bf Acknowledgements}

We thank A. M. Poskanzer and S. A. Voloshin for detailed explanations 
concerning the NA49 flow analysis and useful comments,
and J.-P. Blaizot for careful reading of the manuscript and 
helpful suggestions.

\end{document}